\documentclass[10pt]{article}

\usepackage{geometry}
\usepackage{latexsym, amsmath, amssymb}
\usepackage{psfrag}
\usepackage{mathrsfs}
\usepackage{bbm}
\usepackage[pdftex]{graphicx}
\usepackage{natbib}
\usepackage{color}

\newcommand{\beqn}{\begin{equation}}
\newcommand{\eeqn}{\end{equation}}
\newcommand{\beqa}{\begin{eqnarray}}
\newcommand{\eeqa}{\end{eqnarray}}
\newcommand{\beqanonum}{\begin{eqnarray*}}
\newcommand{\eeqanonum}{\end{eqnarray*}}
\newcommand{\beqnonum}{\begin{equation*}}
\newcommand{\eeqnonum}{\end{equation*}}
\newcommand{\n}{\nonumber}
\newcommand{\jump}{\vspace{0.5cm}}
\newcommand{\bbf}{\begin{bf}}
\newcommand{\ebf}{\end{bf}}
\newcommand{\eqnref}[1]{(\ref{#1})}

\newcommand{\second}{\ensuremath{\mathrm{s}}}
\newcommand{\kg}{\ensuremath{\mathrm{kg}}}
\newcommand{\meter}{\ensuremath{\mathrm{m}}}
\newcommand{\Kelvin}{\ensuremath{\mathrm{K}}}

\newcommand{\km}{\ensuremath{\mathrm{km}}}

\newcommand{\micron}{\ensuremath{\mu\mathrm{m}}}
\newcommand{\Wmsq}{\ensuremath{\mathrm{W/m^2}}}
\newcommand{\WmsqK}{\ensuremath{\mathrm{W/m^2/K}}}

\newcommand{\Kinverse}{\ensuremath{\mathrm{K^{-1}}}}

\newcommand{\ddz}{\ensuremath{\frac{d}{dz}}}

\newcommand{\cotwo}{\ensuremath{\mathrm{CO_2}}}

\newcommand{\Ta}{\ensuremath{T_{\mathrm{a}}}}
\newcommand{\Ts}{\ensuremath{T_{\mathrm{s}}}}
\newcommand{\Tem}{\ensuremath{T_{\mathrm{em}}}}
\newcommand{\pem}{\ensuremath{p_{\mathrm{em}}}}
\newcommand{\zem}{\ensuremath{z_{\mathrm{em}}}}
\newcommand{\gammaav}{\ensuremath{\overline{\Gamma}_m}}
\newcommand{\OLR}{\ensuremath{\mathrm{OLR}}}

\newcommand{\ml}{\mathrm{ml}}
\newcommand{\hml}{\ensuremath{h_\ml}}
\newcommand{\Tml}{\ensuremath{T'_\ml}}
\newcommand{\tauml}{\ensuremath{\tau_\ml}}
\newcommand{\hd}{\ensuremath{h_{D}}}
\newcommand{\Td}{\ensuremath{T'_{D}}}
\newcommand{\taud}{\ensuremath{\tau_{D}}}
\newcommand{\Fdouble}{\ensuremath{F_{\mathrm{2x}}}}
\newcommand{\rhow}{\ensuremath{\rho_{w}}}
\newcommand{\Cw}{\ensuremath{C_{w}}}
\newcommand{\ECS}{\ensuremath{\text{ECS}}}
\newcommand{\TCR}{\ensuremath{\text{TCR}}}

\newcommand{\rhov}{\ensuremath{\rho_{v}}}
\newcommand{\RH}{\ensuremath{\mathrm{RH}}}
\newcommand{\taulambda}{{\ensuremath{\tau_\lambda}}}
\newcommand{\Flambda}{{\ensuremath{F_\lambda}}}
\newcommand{\Ttp}{\ensuremath{T_{\mathrm{tp}}}}
\newcommand{\pvinf}{\ensuremath{p_v^\mathrm{ref}}}

\newcommand{\ppt}{\ensuremath{\partial_T}}

\newcommand{\RHbl}{\ensuremath{\mathrm{RH_{bl}}}}
\newcommand{\rhovstar}{\ensuremath{\rho^*_{v}}}
\newcommand{\Cd}{\ensuremath{C_d}}
\newcommand{\speed}{\ensuremath{u}}
\newcommand{\zref}{\ensuremath{z_{\mathrm{ref}}}}

\newcommand{\RHft}{\ensuremath{\mathrm{RH_{ft}}}}
\newcommand{\qv}{\ensuremath{q_{v}}}
\newcommand{\qvstar}{\ensuremath{q^*_{v}}}
\newcommand{\pvstar}{\ensuremath{p^*_{v}}}

\newcommand{\figurepath}{}

\begin{document}

\title{ The physics of climate change: simple models in climate science}

\author{Nadir Jeevanjee\footnote{Department of Geosciences, Princeton University, Princeton NJ 08544 USA. nadirj@princeton.edu } \footnote{Princeton Program in Atmosphere and Ocean Sciences, Princeton University, Princeton NJ 08540 USA} \footnote{Geophysical Fluid Dynamics Laboratory,  Princeton NJ  08540 USA}
}

\maketitle

\begin{abstract} 
There is a perception that climate science can only be approached with complex computer simulations. But working climate scientists often use simple models to understand their simulations and make order-of-magnitude estimates. This article presents some of these simple models with the goal of making climate science more accessible and comprehensible. 
\end{abstract}

\tableofcontents

\pagebreak

\section{Introduction}
Climate science is, at its roots, a branch of physics. Or, rather, it is an application of several branches of physics: the fluid dynamics of the atmosphere and ocean, the thermodynamics of ideal gases and phase transformations of water, the radiative transfer of sunlight and thermal infrared light, just to name a few. But, too often the connection between the basic physics we know and the climate phenomenon we simulate and observe is lost. The climate is a complex system, exhibiting myriad emergent phenomena which arise from an intricate interplay of the various branches of physics alluded to above,  so understanding often seems out of reach. We thus put our questions about climate change to black-box computer models, but feel uneasy about the results: different models often given quite different answers to the same question, and each model is a patchwork of  fundamental physical laws, inspired guesswork, and parameter `tuning' whose stitches are not visible from the outside.

At the same time, there is \emph{some} simplicity amid the complexity;\footnote{Phrase borrowed from \cite{held2014}.} many climate and climate change phenomena are simulated robustly across models, and some of these phenomena do admit description by simple models which are  grounded in basic physics but are also realistic enough for ballpark estimates. Such models are of great utility, as they give us a basis for reasoning quantitatively about climate change.

Despite this utility, however, many such models have not yet made it into the textbooks or popular literature, but remain scattered throughout the vast and technical scientific literature. Our aim here is to provide a self-contained treatment of a handful of these  models which seem particularly useful. The intended audience are those with a basic math and physics background,\footnote{Introductory sequences in college physics and calculus should suffice.} and in particular scientists and engineers in other fields, who  seek a basis for reasoning for themselves about climate change. 

Given the complexity of the climate system, the simplifications required to make our models tractable will be drastic at times. In this sense we will `lie to tell the truth', but we will endeavor to make our lies explicit, pointing out where approximations are made and where further work is needed (footnotes will be liberally employed to this effect). This article is meant to be pedagogical, rather than a review, so references to the literature are representative rather than exhaustive.  Steps are omitted from some calculations for the sake of concision, but we encourage the reader to fill these in.

We will often focus on the Earth's tropics, sometimes treating it as a stand-in for the entire planet. This is justified to some degree as the tropics account for half the Earth's surface and the majority of it its incoming and outgoing radiation and precipitation. We will also focus on the vertical transports of energy and moisture, and ignore horizontal transports; the latter are crucial for determining the atmospheric circulation and meridional temperature gradients, but the former is perhaps most crucial for determining climate overall, and in particular for determining  the surface temperature \Ts, the central  variable  in climate science. Indeed,  much of the field is focused on understanding how \Ts\ responds to forcing, and how forcing-induced changes in \Ts\ affect other variables such as precipitation, clouds, or humidity. We thus begin by considering  the present day, global and annual mean \Ts\ of 288 K. Where does this number come from?


\section{The atmosphere in radiative-convective equilibrium}
\label{sec_rce}
To understand \Ts\  we will build the simplest possible climate model we can, while at the same time incorporating the physics we know to be essential.  We begin with sunlight. The solar flux at Earth's orbit is  $S_0 = 1360\ \Wmsq$, and this flux is incident on an effective surface area $\pi R^2_E$ (the projected area of the Earth onto a plane perpendicular to the Sun's rays, where $R_E$ is Earth's radius). A fraction $\alpha\approx 0.30$, known as Earth's \emph{albedo}, of this flux is reflected back to space.\footnote{This reflection is primarily  off of clouds and the atmosphere, with the rest contributed by surface reflection, particularly off bright surfaces like desert and ice  \citep{stephens2015, trenberth2009a}. Albedo is thus internally determined by the climate system, but we know of no simple models for cloud albedo, and so here we take the albedo as input to our simple model, rather than predicting it. Simple models do exist, however, for ice albedo; see \cite{budyko1969,sellers1969}, as well as the review in \cite{north1981}.} Dividing by the Earth's total surface area $4\pi R_E^2$ then gives the globally averaged net incoming solar radiation 
\beqn
	S \ \equiv \ \frac{S_0 (1-\alpha)}{4} \ \approx \ 240\ \Wmsq \ .
\eeqn

The most basic physical constraint on climate is that of \emph{planetary energy balance}, which says that $S$ must be balanced by outgoing thermal radiation, also known as the `outgoing longwave radiation', or \OLR. We estimate \OLR\  as blackbody emission, 
\beqn
\OLR = \sigma T^4,
\label{sb_olr}
\eeqn
 for some `emission temperature' \Tem. Our planetary energy balance thus reads
\beqn
	S = \sigma \Tem^4 \ .
	\label{peb}
\eeqn
The observed value of $S$ yields $\Tem = 255$ K.\footnote{By \emph{Wien's law} \citep{gasiorowicz2003},  thermal radiation at $\Tem = 255$ K peaks at infrared wavelengths of order 10 \micron, distinctly longer than the wavelength of solar radiation and visible light; hence the name `longwave'.} This is much colder than the observed global average \Ts\ of 288 K, but is a reasonable estimate of the vertically-averaged \emph{atmospheric} temperatures, consistent with the fact that OLR largely emanates not from the surface, but from greenhouse gases\footnote{Note that greenhouse gases are simply gases that are able to emit and absorb thermal infrared radiation.} in the atmosphere (largely water vapor and carbon dioxide).  But given the atmospheric \Tem, how can we find \Ts? How are surface and atmospheric temperatures related? The planetary energy balance \eqnref{peb} cannot help us here;  additional physics is required. 


\subsection{Heuristics of RCE}
\begin{figure}[t]
        \begin{center}
                \includegraphics[scale=0.5,trim = 0cm 1cm 0cm 2.5cm, clip=true]{\figurepath 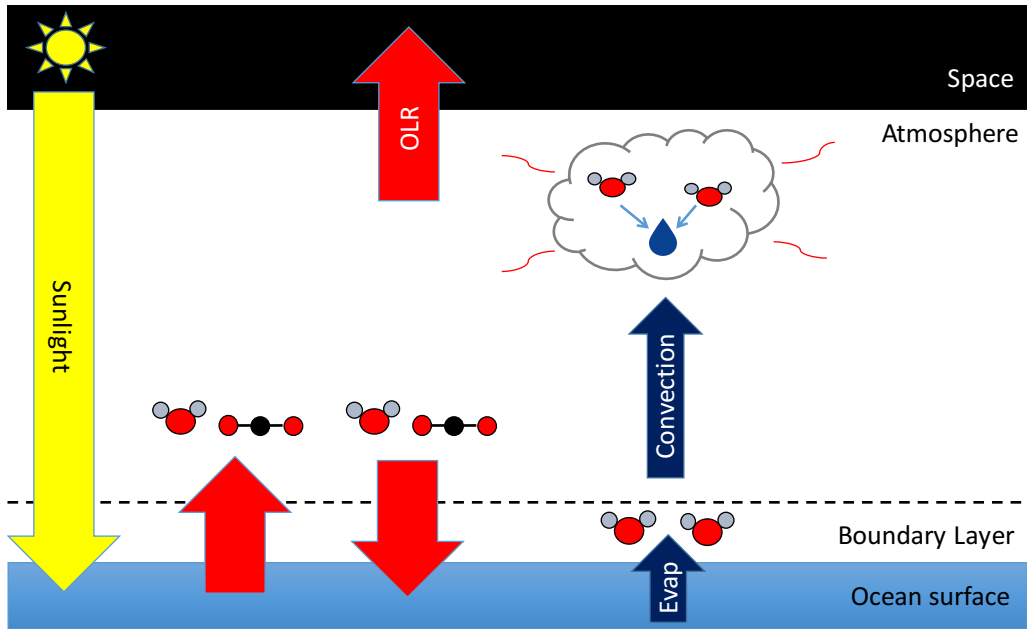}
                \caption{A cartoon of radiative-convective equilibrium. See text for explanation.
                \label{rce_cartoon}
                }
        \end{center}
\end{figure}

To find this additional physics, we need a picture for how energy flows through the Earth system; a highly simplified such picture is given in Fig. \ref{rce_cartoon}.\footnote{See \cite{trenberth2009a} for a more comprehensive picture.} Sunlight heats the ocean, but the ocean cannot cool itself by radiating this heat away, because the greenhouse gases (particularly water vapor) in the lower atmosphere simply radiate most of this heat back to the ocean (Fig. \ref{rce_cartoon}, opposing red arrows). Facing such opposition, the ocean ends up cooling the same way we humans do: by evaporation. Evaporation occurs into the `boundary layer' (or BL), the bottom kilometer or so of atmosphere which is in close contact with the surface and in which there are no clouds. Eventually this water vapor is transported up and out of the boundary layer  by convection. As this convecting air rises and cools (see Section \ref{sec_simpleRCE} below), water vapor condenses, resulting in cloud and rain formation. This condensation heats the atmosphere, via the release of the same latent heat which was required as an input for evaporation. This heat from condensation, released at altitude where the greenhouse effect is weak, can now radiate out to space unimpeded. Thus, the planetary energy balance is not purely radiative but is mediated by convection, with water vapor as the key middleman. The atmosphere is thus in a state of  \emph{radiative-convective equilibrium}, or RCE for short. 

The part of the atmosphere in RCE in which this  mediation by convection (and other forms of weather in the extratropics) occurs is known as the \emph{troposphere}.\footnote{`tropos' = turning, referring to the overturning of air masses by convection and other weather phenomenon.} The troposphere occupies the bottom 15 km of the atmosphere in the tropics and roughly 9/10 of the atmosphere's mass (i.e. extends up to pressures of about 0.1 atm), and is really what is pictured in Fig. \ref{rce_cartoon}. Above the troposphere lies the stratosphere and other air masses, which are heated primarily by ultraviolet solar absorption rather than convection, and are thus in pure \emph{radiative equilibrium} rather than RCE. We focus on the troposphere in what follows.

\subsection{A simple RCE climate model}
\label{sec_simpleRCE}
Given this picture, the simple climate model we seek must be a model of the troposphere in RCE.\footnote{Rather than of pure radiative equilibrium, as sometimes put forth in introductory texts \citep{hartmann2015book,vallis2012,randall2012}.}  Thus we must  somehow use the fact that convection heats the troposphere to relate the surface temperature to tropospheric temperatures. To proceed we need two facts:
\begin{enumerate}
	\item The temperature of a convecting parcel of air decreases as it rises, at a characteristic rate of roughly 7 K/km.
			\label{lapse}
	\item To a good approximation, the temperature profile of the troposphere is simply that of a rising convecting parcel.
			\label{zerob}
\end{enumerate}
 Fact \ref{lapse} can be understood as follows. Consider a rising convective parcel that conserves its mass and rises rapidly such that heat exchange with the environment is negligible. This process is governed by the first law of thermodynamics
	\beqn
		dQ = dU + p\, dV
		\label{firstlaw}
	\eeqn
	where $dQ$ is the heat gained by the system, $U$ is its internal energy, $p$ is its pressure and $V$ its volume. For an ideal gas we have $U= \rho V C_vT$ as well as the ideal gas law 
	\beqn
	 p = \rho R_dT
	\label{ideal_gas}
	 \eeqn
	  (where $\rho$ is the parcel density in $\kg/\meter^3$, $C_v $ the specific heat capacity at constant volume, and $R_d$ the specific gas constant for dry air, with the latter two in  J/kg/K). Assuming an adiabatic process ($dQ=0$),  we rearrange \eqnref{firstlaw} into
	\beqn
		0 = \rho C_p\,dT - dp
		\label{adiabatic}
	\eeqn
	 where the specific heat at constant pressure $C_p = C_v +R_d$. Further assuming hydrostatic balance\footnote{There is some sleight-of-hand here, as hydrostatic balance typically applies to the pressure and density of the quiescent \emph{environment}, rather than a rising parcel. For most applications (such as ours) this approximation is permissible, but if we take  it to extremes (such as applying the dry adiabatic lapse rate \eqnref{dry_lapse} over large enough distances to generate negative temperatures)  it can lead to nonsensical results; see \cite{romps2015mse} for further discussion.}
	 \beqn
		  dp/dz = -\rho g
	 \label{hyd_balance}
	  \eeqn
	   and rearranging yields the `dry adiabatic lapse rate'
	 \beqn
	 	\Gamma_d \equiv -\frac{dT}{dz} = - g/C_p\ .
		\label{dry_lapse}
	\eeqn
	Evaluating this for $g = 10\ \mathrm{m/s^2}$ and $C_p = 1000$ J/kg/K tell us that a dry air parcel cools at a rate of $\Gamma_d = 10$ K/km. Thus, even though hot air rises, it \emph{cools} as it rises from adiabatic expansion, just like the cool air escaping from the valve of a pressurized bicycle tire. If we add in the effects of water vapor, which condenses as the parcel rises, releasing latent heat and thus reducing the parcel's cooling rate, we find that the `moist adiabatic lapse rate' $\Gamma_m$ is no longer constant in the vertical, but varies from about 4 K/km near the surface to the dry value of 10 K/km in the upper troposphere, with an average  value of  $\gammaav \approx  7$ K/km, as claimed above (a formula for $\Gamma_m$ is given in  Eqn. \eqnref{gamma_m} in Section \ref{sec_RHft}). 
	
	As for Fact \ref{zerob} above, it is found in both observations and simulations \citep[e.g.][]{mapes2001,held1993} that the actual temperature profile $\Ta(z)$ of the tropical troposphere  satisfies $-d\Ta/dz \approx \Gamma_m$. This can be understood as a kind of equilibrium. If $-d\Ta/dz < \Gamma_m$, then a moist rising parcel will cool relative to its environment and become negatively buoyant and hence cease rising. Such an atmosphere won't convect, and will cool radiatively (via infrared emission to space from greenhouse gases) until $-d\Ta/dz \approx \Gamma_m$.  If on the other hand $-d\Ta/dz > \Gamma_m$, then by the converse to the previous argument such an atmosphere is conducive to moist convection, and convection will  warm the atmosphere until $-d\Ta/dz \approx \Gamma_m$. 
	
Combining facts 1 and 2, and assuming that parcels near the surface have temperatures approximately equal to the surface temperature \Ts, we then have the following relation between tropospheric and surface temperatures:
\beqn
	\Ta(z) \ =\  \Ts - \gammaav z \ .
	\label{eqn_Ta}
\eeqn
This tells us, rather remarkably, that convection pegs the tropospheric temperature profile $\Ta(z)$ to \Ts, and that the two cannot be varied independently.  This stands in contrast to non-convecting regions such as the arctic, where \Ts\ and \Ta\ in fact \emph{may} vary independently \citep{payne2015}.

Equation \eqnref{eqn_Ta} is the other essential  ingredient required to complete our climate model and determine \Ts.  The final step is to note that radiative cooling occurs rather uniformly throughout the troposphere \citep[e.g.][]{hartmann2015book}, and so  we may assume that \Tem\ represents a vertically-averaged tropospheric temperature, and should thus occur around halfway through the troposphere at an `emission pressure' of $\pem = 0.5$ atm. A quick calculation\footnote{Assume an isothermal atmosphere with $T=\Tem$, which from \eqnref{hyd_balance} and \eqnref{ideal_gas} yields $p = (1\ \text{atm}) \exp(-z/H)$, where $H\equiv R_d\Tem/g \approx 7.5\ \km$ is the atmospheric `scale height'. Setting $\pem = 0.5$ atm then yields $\zem = H \ln2 \approx 5\ \km$.} then yields a  corresponding emission height of  $\zem\approx 5$ km. Substituting this into  \eqnref{eqn_Ta} and rearranging then yields  
\beqn
	\Ts \ = \ \Tem + \gammaav \zem\  = \ 290 \ \Kelvin \ .
	\label{eqn_Ts}
\eeqn     	
This is remarkably close to the actual value of 288 K. Furthermore,  Eqn. \eqnref{eqn_Ts} tells us that this value stems from the radiative energy balance of the planet (as embodied in \Tem), combined with the effects of convection on the temperature profile of the troposphere (as embodied in \gammaav).


\section{A two-box model for transient and equilibrium climate sensitivities}
\label{sec_2box}
Now that we have a picture of Earth's energy flows (Fig. \ref{rce_cartoon}) and a corresponding simple model of RCE [Eqns. \eqnref{peb} and \eqnref{eqn_Ts}], we can begin to think about climate change. We will take a transient point of view, asking in what order and on what timescales various components of the Earth system change, with an emphasis on the oceanic response for which we will develop a simple quantitative model.

Global climate model (GCM) calculations show that a doubling of \cotwo\ instantaneously decreases the \OLR\ by $\Fdouble \approx 3.6 \ \Wmsq$ \citep{myhre1998,wilson2012}.
 In our crude RCE model, this can be thought of as an instantaneous increase in \zem, which lowers \Tem\ and hence  the OLR,  breaking the planetary energy balance \eqnref{peb} and causing energy to accumulate in the system. But how, specifically, do we expect the system to respond? Given the heuristics of Fig. \ref{rce_cartoon}, we expect the following:
\jump

\begin{tabular}{p{3cm} p{1cm} p{10cm}}
OLR decreases & $\implies$ & 	Precipitation decreases (because condensation heating balances radiative emission from the troposphere) \\
 & & \\
 & $\implies$ & BL moistens (because the rate at which convection converts BL moisture to precipitation has decreased)  \\
 & & \\
 & $\implies$ & Evaporation decreases (because evaporation is inhibited by humidity, cf. \eqnref{bulk_E1} in Section \ref{sec_rh}) \\
 & & \\ 
 & $\implies$ & \Ts\ increases (because evaporation is how the ocean cools itself) \\
\end{tabular}
\jump

What are the associated timescales for these processes? For atmospheric processes such as BL moistening, the timescale is roughly a day.\footnote{This can be seen using the bulk aerodynamic formulae for evaporation \eqnref{bulk_E1} which we present in Section \ref{sec_rh}. From this equation and the fact that the water mass (in $\kg/\meter^2$) in a boundary layer of height $h$ is $\rhov h$ (neglecting variations of \rhov\ with height), one finds that perturbations to \rhov\ in the BL decay with a characteristic timescale of $h/(\Cd \speed) \approx (500\ \meter)/(10^{-3} \times 5\ \meter/\second) \approx \text{1 day}$.} But what about the increase in \Ts?

\begin{figure}[t]
        \begin{center}
                \includegraphics[scale=0.5,trim = 0cm 1.5cm 0cm 5cm,clip=true]{\figurepath 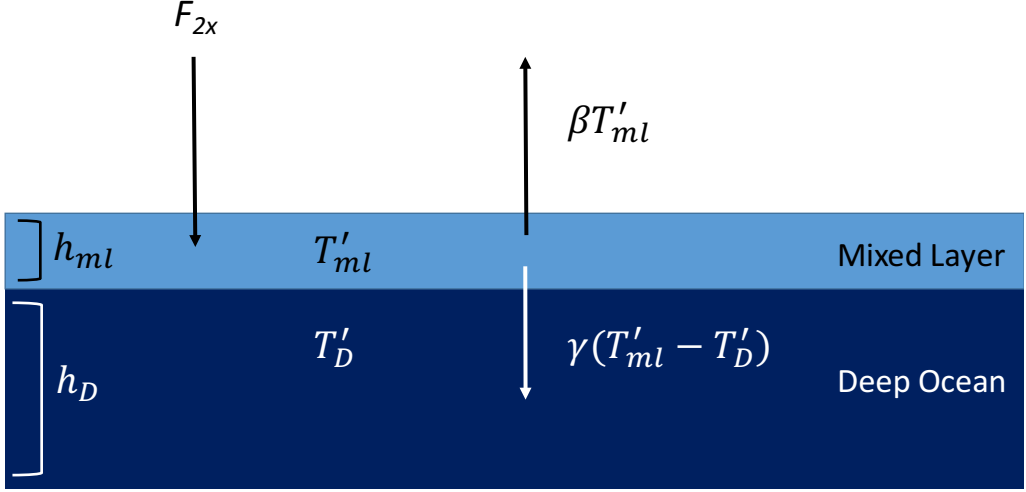}
                \caption{Two-box model for the ocean. See text for discussion.
                \label{fig_2box}
                }
        \end{center}
\end{figure}

To answer this, we need to know something about the heat capacity of the ocean. In general, the ocean is comprised of its own oceanic boundary (or mixed) layer, with depth $\hml \approx  100$ m, and a deep ocean with a global average depth\footnote{Obtained by taking the average ocean depth of 4000 m and multiplying by 2/3, the fraction of the globe covered by ocean.} of $\hd =  2500$ m, over an order of magnitude greater.  As the mixed layer warms in response to the decreased evaporation, it will develop a uniform temperature anomaly \Tml\ (which is also the \Ts\ anomaly) which will spread to the atmosphere via \eqnref{eqn_Ta}, causing an increase in the net top-of-atmosphere radiation $\OLR - S$. We linearize this increase and write it as
 $\beta\Tml$, where
\beqn
	\beta \ \equiv\  \frac{d\, \OLR}{d \Ts}  \ - \  \frac{dS}{d\Ts} \ .
	\label{beta_def}
\eeqn
 (These derivatives are assumed to be taken at the \emph{fixed}, doubled \cotwo\ concentration, and tell you how much \Ts\ needs to increase until \OLR\ is back in balance with $S$.) The \Tml\ anomaly will also cause an increased heat export to the deep ocean, which we similarly linearize as  $\gamma(\Tml-\Td)$. These fluxes are depicted in Fig. \ref{fig_2box}. Typical values, obtained from coupled atmosphere-ocean GCMs, are $\beta \approx \gamma \approx 1\ \Wmsq/\Kelvin$ \citep[][]{geoffroy2013,dufresne2008}. An order-of-magnitude estimate for $\beta$ can be obtained by neglecting $dS/\Ts$ and appealing to our simple RCE climate model \eqnref{sb_olr} and \eqnref{eqn_Ts}:  
\beqn
\beta_{\text{blackbody}} \ =\ \frac{d \OLR}{d \Tem} \,\frac{d \Tem}{d \Ts} \ = \  \ 4\sigma \Tem^3  \ \approx  \ 3.5 \ \Wmsq/\Kelvin\ .
\label{beta_bb}
\eeqn
 While the order-of-magnitude is right, this is also a significant over-estimate, a point we'll return to later when we discuss the water vapor feedback. As for $\gamma$, there do not seem to be any simple models, even for just order-of-magnitude estimation.

Setting \rhow\ and \Cw\ as the densities and specific heat capacities of water, we then have the following two box model for the mixed layer and deep ocean response to \cotwo\ forcing \citep[e.g.][]{geoffroy2013, vallis2012, held2010}:
\begin{subequations}
	\beqa
		  \rhow\Cw \hml \frac{d\Tml}{dt}& = & \Fdouble - \beta \Tml - \gamma(\Tml-\Td) \label{eqn_ml} \\ 
	  	\rhow\Cw \hd \frac{d\Td}{dt}  & = &  \gamma(\Tml-\Td) \label{eqn_deep}
 	\eeqa
	\label{eqns_2box}
 \end{subequations}
 
The much larger depth and hence heat capacity of the  deep ocean suggests that it will take much longer to respond to the \Fdouble\ forcing than the mixed layer. What are these respective timescales? If we assume that the deep ocean hasn't responded yet,  (i.e. if we fix $\Td \equiv 0$), then \eqnref{eqn_ml} tells us that the mixed layer warms with a characteristic timescale
 \beqn
 \tauml = \frac{\rhow\hml\Cw}{\gamma + \beta} \approx \mbox{6 years}\ .
\eeqn
 If we now consider timescales longer than \tauml\ then we may set $d\Tml/dt =0$ in \eqnref{eqn_ml}, solve for \Tml, and plug that into \eqnref{eqn_deep}. The resulting equation has a characteristic timescale of
 \beqn
 \taud = \rhow\hd\Cw\frac{\gamma + \beta}{\gamma \beta} \approx \mbox{600 years .}
\eeqn
Thus the vast difference in total heat capacity between the mixed layer and deep ocean indeed leads to a separation of timescales in their responses to forcing, and thus to two timescales for global warming: a `fast' timescale of about $\tauml \approx 6$ years during which the mixed layer equilibrates (really a `quasi-equilibrium', since this equilibrium state will change as the deep ocean response), and a `slow'  timescale of $\taud\approx 600$ years during which the deep ocean equilibrates (a true equilibrium).   On intermediate timescales in between \tauml\ and \taud\ in which the mixed layer is in quasi-equilibrium, we can assume $d\Tml/dt = 0$ and $\Td \approx 0$, which from \eqnref{eqn_ml} yields 
\beqn
	\Tml \ = \ \frac{\Fdouble}{\gamma + \beta}\ \approx\ 1.8 \ \Kelvin \ .
	\label{eqn_TCR}
\eeqn 
 This is essentially the \emph{transient climate response} (TCR), which measures the climate response to a doubling of \cotwo\  before the deep ocean has responded. TCR should be contrasted with the \emph{equilibrium climate sensitivity} (ECS), which is the surface warming which occurs after \emph{both} the mixed layer and deep ocean have reached a mutual equilibrium ($d\Tml/dt = d\Td/dt =  0$), after several hundred years. In our two-box model \eqnref{eqns_2box} the ECS is
 \beqn
 	\Tml \ = \ \frac{\Fdouble}{\beta}\ \approx \  3.6\ \Kelvin \ ,
	\label{eqn_ECS}
\eeqn 
 or about twice the TCR. Note that $\text{ECS} \sim 1/\beta$. The ECS can be thought of as a kind of `committed' warming for a given \cotwo\ concentration,\footnote{It is very important to note, however, that experiments with a \emph{fixed} \cotwo\ concentration are highly idealized in that they neglect the \emph{carbon cycle}, i.e. the fact that \cotwo\ emissions do not  all stay in the atmosphere but are also partitioned into the ocean, land, and eventually the deep earth \citep[e.g.][]{depaolo2015,archer2008}. More realistic simulations with prescribed \cotwo\ \emph{emissions} (rather than concentrations) and an interactive carbon cycle find that the eventual, committed warming is a function not of \cotwo\ concentrations but rather \emph{cumulative emissions} \citep[e.g.][]{allen2009,matthews2009}.} and since $\gamma > 0$, Eqns. \eqnref{eqn_TCR} and \eqnref{eqn_ECS} imply 
 \beqn
	 \ECS\  >\   \TCR \ .
 \eeqn
 This is because on the intermediate timescales during which $\Tml = \TCR$, the mixed layer is both radiating heat to space \emph{and} exporting heat to the deep ocean, and can thus come to (quasi-)equilibrium at a lower temperature.

\section{The water vapor feedback}
\label{sec_h2o_feedback}
In the last section we introduced the quantity $\beta$, which (when we neglect $dS/d\Ts$) is just $d \OLR/ d\Ts$. We quoted a value of $\beta$ from GCMs of $1\ \Wmsq/\Kelvin$, but our simple RCE model yields a value of roughly $3.5\ \Wmsq/\Kelvin$, almost a factor of four off. What physics is our RCE model missing?

Our model \eqnref{sb_olr} assumes that the atmosphere emits as a blackbody, or an object at uniform temperature that absorbs and emits\footnote{A basic law of physics, \emph{Kirchoff's law}, states that the absorptivity and emissivity of an object are equal, so a strong absorber is necessarily a strong emitter, and a weak absorber is a weak emitter.} perfectly at all wavelengths. This assumption fails on multiple counts, however.  As manifest in Eqn. \eqnref{eqn_Ta}, the troposphere does not have a uniform temperature, but rather exhibits temperatures ranging roughly from 200 - 300 K.  Furthermore, the atmosphere is clearly not a perfect absorber at all wavelengths; this is obvious in the visible part of the radiation spectrum,  and is also true in some regions of  the thermal infrared. Indeed, individual greenhouses gases such as carbon dioxide and water vapor absorb and emit preferentially at some wavelengths and less so at others, yielding characteristic emission levels  $\zem(\lambda)$ and temperatures $\Tem(\lambda)$ which depend on wavelength $\lambda$. The key ingredient in building a more refined model of \OLR\ is 
understanding $\Tem(\lambda)$, because then the \OLR\ may be estimated\footnote{We are here employing the unit optical depth approximation discussed below.}
 using the spectrally resolved Planck density\footnote{The Planck density can be derived from first principles \citep[e.g.][]{kittel1980}, and with wavelength as the spectral coordinate  takes the form
\beqn
B(\lambda,T) \ = \ \frac{2hc^2}{\lambda^5}\frac{1}{\exp\left(\frac{hc}{\lambda k_b T}\right)-1}\ .
\eeqn
Here $h$ is Planck's constant, $k_b$ is Boltzmann's constant, and $c$ is the speed of light. The Planck density satisfies $\int_0^\infty \pi B(\lambda,T) \, d\lambda \ = \ \sigma T^4$ \citep{gasiorowicz2003}, and is thus a spectral refinement of \eqnref{sb_olr}.}
 $B(\lambda,T)$ (units $\Wmsq/\mathrm{sr/m}$) as 
\beqn
	\OLR = \int \pi B(\lambda,\Tem(\lambda)) \, d\lambda \ .
	\label{planck_olr}
\eeqn
(Here the factor of $\pi$ accounts for integration over solid angle, yielding an \OLR\ flux in \Wmsq.) We thus turn to the question of what sets the emission level at a given $\lambda$, focusing on water vapor emission since water vapor absorbs and emits effectively across a much wider  range of $\lambda$  than any other greenhouse gas.	


\subsection{Emission from unit optical depth}
\begin{figure}[t]
        \begin{center}
        		\begin{tabular}{ccc}
                 \includegraphics[scale=0.4,trim=1cm 4cm 13cm 2cm, clip=true]{\figurepath 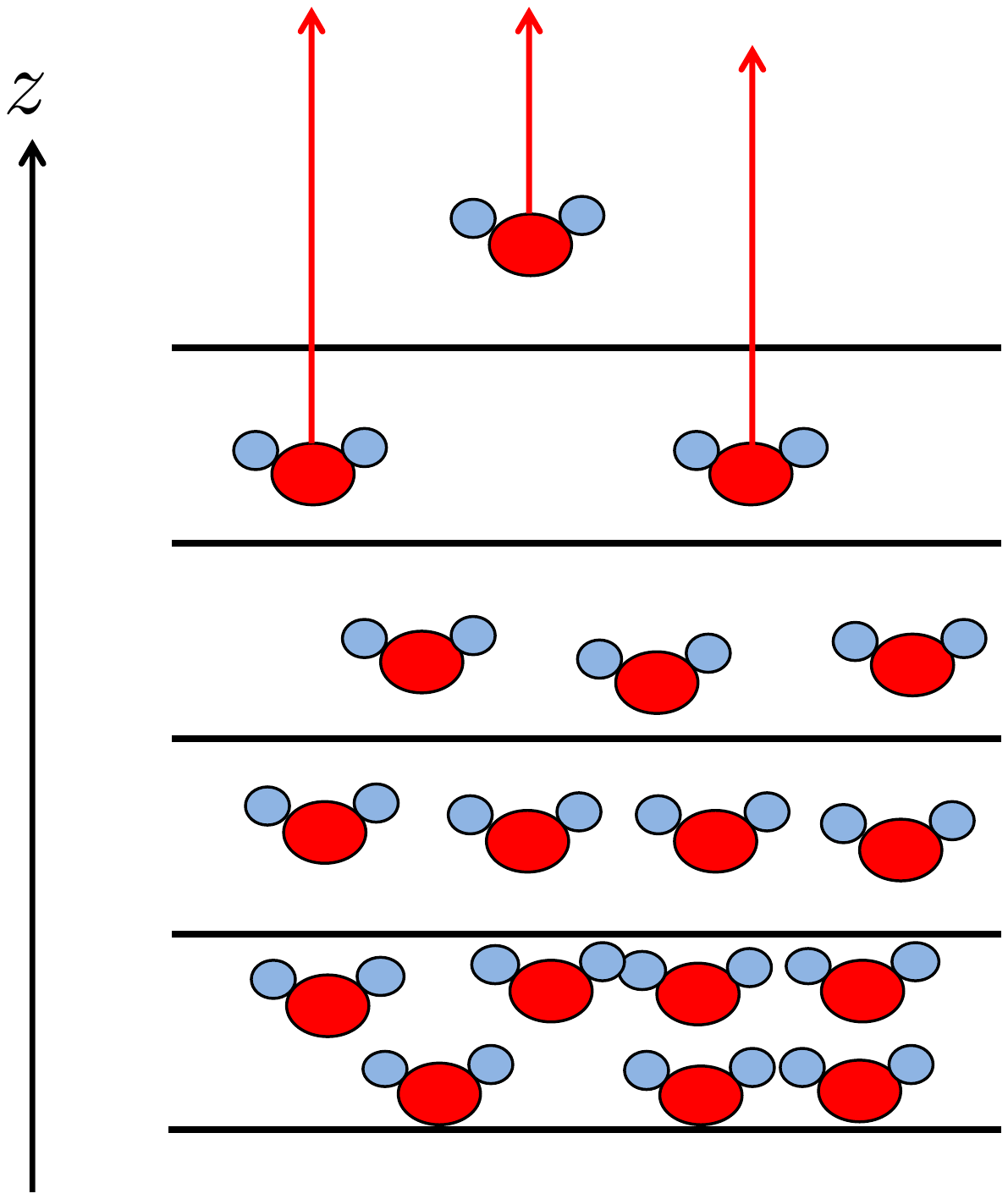} &
                  \includegraphics[scale=0.4,trim=4cm 4cm 13cm 2cm, clip=true]{\figurepath 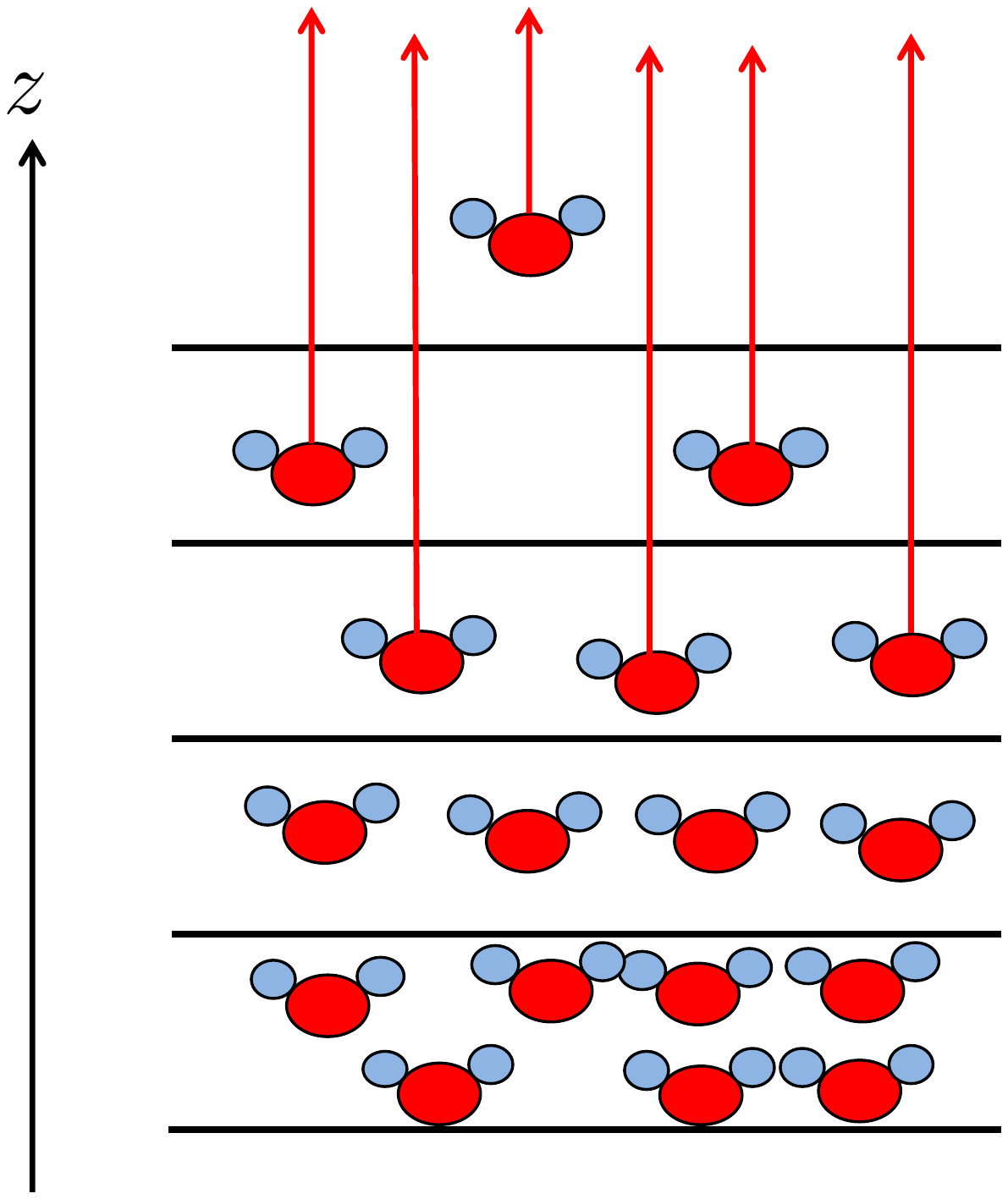} &
                  \includegraphics[scale=0.4,trim=4cm 4cm 10cm 2cm, clip=true]{\figurepath 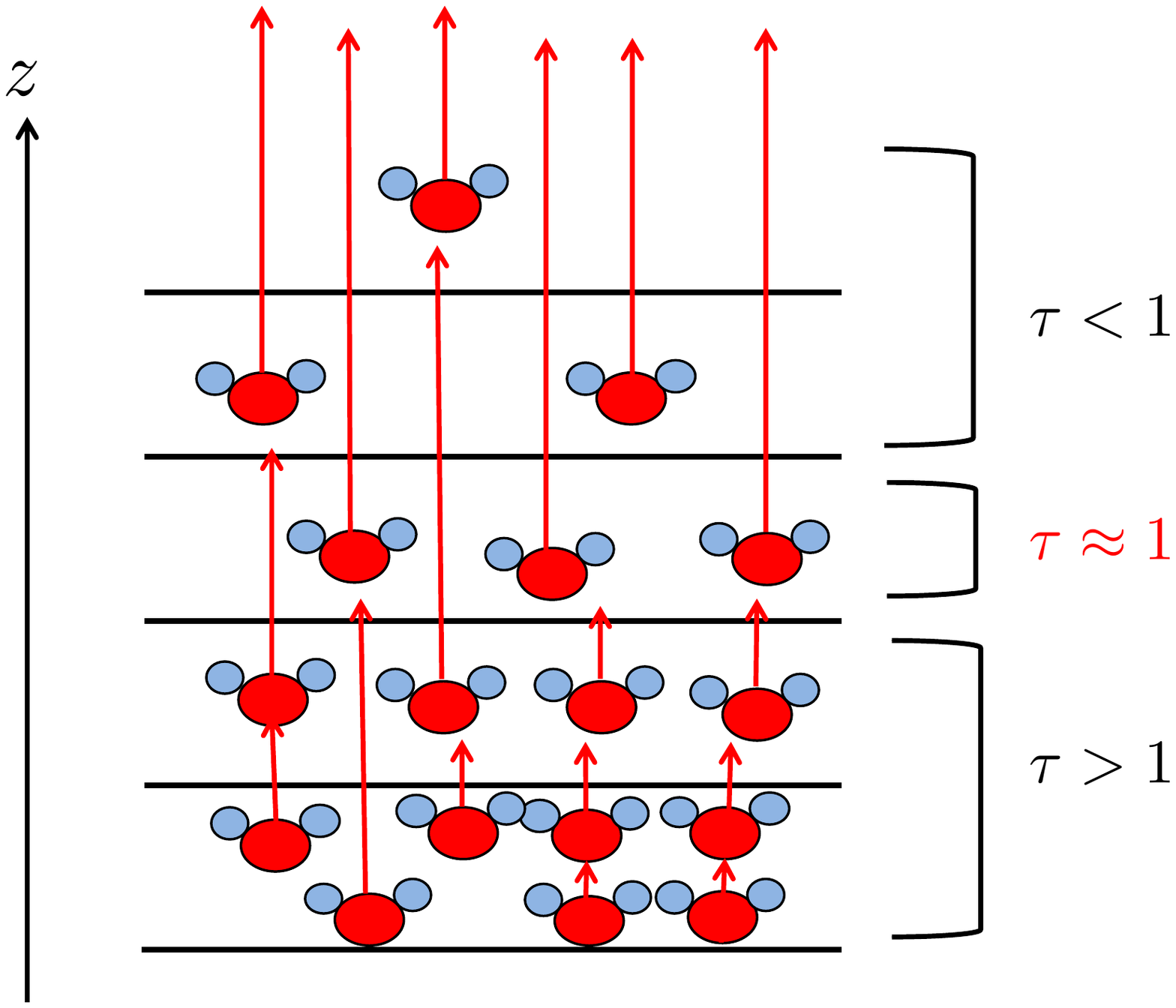} 
             \end{tabular}                   
              \caption{Cartoon of emission to space, which maximizes at a sweet spot where the optical depth $\tau \approx 1$.
              \label{tau_cartoon}
               }
        \end{center}
\end{figure}

Consider an atmospheric column with water vapor molecules whose density \rhov\ decreases exponentially with height (cf. \eqnref{eqn_rhov} below), and let us consider the  emission to space (i.e. the contribution to the \OLR) from these molecules, as pictured in Fig. \ref{tau_cartoon}. The top two layers in Fig. \ref{tau_cartoon} have little difficulty emitting to space because their view is unobstructed, but the density of emitters in these layers is relatively low, so the emission will also be low. In the third layer, the molecules' view of space is still unobstructed (just barely), and the density is higher, so emission to space is higher. For layers four and five there are plenty of emitters, but their view is almost totally obstructed, so their emission to space is very low. Thus, emission to space is maximized around a `sweet spot' where the absorbers/emitters above have not yet totally obstructed the view of space, but the density is high enough for emission to be appreciable. This sweet spot will be our emission level.

To formalize this we need the notion of \emph{optical depth}, defined as\footnote{We neglect here the \emph{pressure-broadening} of absorption coefficients; this was argued by \cite{ingram2010} not to affect the conclusions presented in this section, but further work on this question is needed.}  \citep{hartmann2015book, pierrehumbert2010}\
\beqn
	\taulambda(z) \ \equiv \ \underbrace{\kappa(\lambda)}_{\mathrm{(m^2/kg)}}\underbrace{\int_z^\infty \rhov \, dz}_{\mathrm{(kg/m^2)}} \ = \ \frac{\text{Effective area of absorbers}}{\text{Actual area of column}} \ .
	\label{eqn_tauz}
\eeqn
Here \rhov\ is the density of water vapor (typically 2\% or less of the dry air density $\rho$), and $\kappa(\lambda)$ gives the effective cross-sectional area of water vapor molecules at wavelength $\lambda$ per unit mass, in $\meter^2/\kg$. The integral of \rhov\ simply gives the mass of water vapor in the column above per unit area ($\kg/\meter^2$), so \taulambda\ can be interpreted as the ratio of the total effective area of absorbers above height $z$ to the actual area of the column, as depicted in \eqnref{eqn_tauz}. Applying this to the cartoon in Fig. \ref{tau_cartoon}, we see that above our sweet spot we have $\taulambda < 1$ (the `optically thin' regime) and below our sweet spot we have $\taulambda > 1$ (the `optically thick' regime), and thus our sweet spot occurs around $\taulambda\approx 1$. For simplicity we further assume that \emph{all} the emission occurs at exactly $\taulambda=1$; we refer to this as the \emph{unit optical depth approximation}, and with it in mind we define our emission height $\zem(\lambda)$ by the relation 
\beqn
	\taulambda(\zem(\lambda)) = 1  .
	\label{eqn_tauz=1}
\eeqn

\subsection{\Ts-invariance of water vapor emission temperatures}
Let us now evaluate \taulambda\ in \eqnref{eqn_tauz}. The key step is to write the vapor density \rhov\  as 
		\beqn
			\rhov \ =\ \frac{p_v}{ R_v T} \ = \ \frac{\RH p_v^*(T)}{R_v T}
			\label{eqn_rhov}
		\eeqn
		where the first equality follows from the ideal gas law and the second is just a definition of the relative humidity \RH\ (note that \RH\ may vary in the vertical). Here $p_v$ is the partial pressure of water vapor, $R_v$ is the specific gas constant for water vapor,  and 
\beqn
	p_v^*(T) \ = \ \pvinf \exp(-L/R_v T)
\label{CC}
\eeqn
 is the Clausius-Clapeyron expression for the saturation partial pressure of water vapor, with reference pressure $\pvinf = 2.5 \times 10^{11}$ Pa and $L$ the latent heat of vaporization (in J/kg). Note that this expression is fundamental to our RCE picture of the atmosphere, as it determines how much the water vapor content of a parcel decreases (and hence how much condensation is produced) per degree of cooling.\footnote{This cooling can be adiabatic, as for the rising parcels of section \ref{sec_rce}, or it can be diabatic, as for nighttime air close to rapidly cooling objects on the ground onto which dew is then deposited. Note also that \eqnref{CC} also explains the dryness of cold winter days, as well as the that fact that it can sometimes be `too cold' to snow appreciably.} 
 
 The key point about Eqn. \eqnref{eqn_rhov} is that, up to vertical variations in \RH\ (which are typically not greater than a factor of two over the entire troposphere), $\rhov = \rhov(T)$, i.e. \rhov\ is only a function of temperature. In other words, given the temperature of an air parcel, regardless of its height, pressure, or the surface temperature below it, Eqn. \eqnref{eqn_rhov} tells us the value of \rhov, at least up to a factor of \RH\ (note that this is still quite meaningful since the exponential in \eqnref{CC} means that \rhov\ varies by roughly four orders of magnitude over the depth of the troposphere).
 
The fact that \rhov\ is essentially  a function of temperature only suggests that we switch coordinates in \eqnref{eqn_tauz} from $z$ to $T$. This yields
\beqn
	\taulambda(T) \ = \ \kappa(\lambda) \int_{\Ttp}^{T}  \rhov(T')\, \frac{dT'}{\Gamma_m} 
	\label{eqn_taut}
\eeqn
where we have again assumed that $-dT/dz$ is given by the moist adiabatic lapse rate $\Gamma_m$. (The lower bound of this integral \Ttp\ is the `tropopause' temperature, i.e. the temperature at the top of the troposphere.\footnote{The tropopause also forms the lower boundary of the stratosphere. In setting \Ttp\ as the lower bound of our integral we are ignoring the small amount of water vapor in the stratosphere.}) Now, although $\Gamma_m$ varies in the vertical, it turns out that $\Gamma_m$ is also essentially a function of temperature alone (we demonstrate this in Section \ref{sec_RHft} below, using Eqn. \eqnref{gamma_m});  thus the entire integrand in \eqnref{eqn_taut} is essentially a function of $T'$ alone. 

We may then ask how Eqn. \eqnref{eqn_taut} changes with surface temperature \Ts. But, we have just argued that $\rhov$ and $\Gamma_m$ are determined locally by their temperature alone, and are thus ignorant of  \Ts!  Thus the integrand in \eqnref{eqn_taut} does not depend on \Ts, and hence neither does $\taulambda(T)$. But this means that  the emission temperature $\Tem(\lambda)$, which by definition satisfies
\beqn
	\taulambda(\Tem(\lambda)) = 1  \ ,
	\label{eqn_taut=1}
\eeqn
is also independent of \Ts! This has major implications for the water vapor feedback, as we will see momentarily. Going forward, we will refer to any quantity which is independent of \Ts\ (such as $\rhov(T)$, $\Gamma_m(T)$, or $\Tem(\lambda)$) as  \emph{\Ts-invariant}.

\subsection{Simpson's paradox and the water vapor feedback}
The argument that $\Tem(\lambda)$ is \Ts-invariant goes back to \cite{simpson1928}, and leads to what is known as \emph{Simpson's paradox}: if $\Tem(\lambda)$ does not depend on \Ts, then by \eqnref{planck_olr} neither does the \OLR!  This is both unrealistic, as \OLR\ is known to vary  significantly across the globe \citep{hartmann2015book}, and is also unphysical, since a single allowed value of \OLR\ would make the climate unstable.\footnote{A scenario known as the `runaway greenhouse' and thought to be relevant to the early history of venus; see \cite{pierrehumbert2010, ingersoll1969}.} Indeed,  a \Ts-independent \OLR\  (and $S$) implies $\beta = 0$ by \eqnref{beta_def}. We have thus perfectly overshot our goal of lowering  our estimate of $\beta$ from  \eqnref{beta_bb}, and moreover we obtain an infinite climate sensitivity by \eqnref{eqn_ECS}. We have clearly erred egregiously somewhere along the way. But where?

The resolution of Simpson's paradox must lie in some of the approximations we have made. It turns out that the most egregious  one is one which was only implicit: namely, we assumed that that for all $\lambda$,  $\taulambda =1$ somewhere in the atmosphere, or equivalently that Eqns. \eqnref{eqn_tauz=1} and \eqnref{eqn_taut=1} possess a solution for all $\lambda$.  In fact, there is a wide swath of the infrared spectrum for which water vapor is optically thin, even at the surface, and hence Eqns. \eqnref{eqn_tauz=1} and \eqnref{eqn_taut=1} have no solutions; this spectral region is known as the \emph{water vapor window}, and is generally considered to be   $8< \lambda < 12\ \micron$ \citep{petty2006}.

Given this, we may proceed to a partly-simpsonian model \citep{ingram2010}, wherein we assume that in the window $\tau(z) \equiv 0$ and contributions to the \OLR\ come directly from the surface, and outside the window $\tau=1$ somewhere in the atmosphere, and hence $\Tem(\lambda)$ exists and is  \Ts-invariant. This implies that 
\beqn
	\OLR \ =\  \int_{8 \ \micron}^{12\ \micron}  \pi B(\lambda,\Ts)\, d\lambda \ + \  \int_{\lambda\, \notin \, (8\, \micron,\, 12\ \micron)} \pi B(\lambda,\Tem(\lambda)) \, d\lambda \ . 
\eeqn
Our partly-Simpsonian estimate of $\beta$ (continuing to assume $dS/d\Ts = 0$) is then
\beqn
	\beta_{\text{ps}} \ =\  \int_{8 \ \micron}^{12\ \micron} \pi \frac{dB(\lambda,\Ts)}{d\Ts} \, d\lambda \ \approx \ 2\ \Wmsq/\Kelvin 
	\label{beta_ps}
\eeqn
where the integral is computed numerically. Thus, taking into account the spectral nature of thermal radiation as well as the \Ts-invariance of water vapor optical depth leads to almost a halving of $\beta$, and thus an almost doubling of climate sensitivity. 
First discovered by \cite{manabe1967}, this has become known as the \emph{water vapor feedback},\footnote{In most of the literature this feedback is described as the effect of allowing specific humidities $\rhov/\rho$ at fixed pressure to change, and is often combined with the \emph{lapse-rate} feedback which describes the effect of vertically non-uniform warming. What we call the water-vapor feedback here is essentially the sum of these two effects, with some minor differences; see \cite{held2012} and \cite{ingram2013a} for further discussion.} and gets us about halfway from our first $\beta_{\text{blackbody}} = 3.5\ \WmsqK$ estimate towards the $\beta \approx 1 \ \WmsqK$ produced by our comprehensive climate models.

What about the other factor of two? There are many potential reasons for this error, including our neglect of the water-vapor `continuum',\footnote{ which provides non-negligible optical depths throughout the water vapor window; see, e.g., \cite{shine2012}.} as well as our neglect of solar radiation feedbacks, i.e. the $d S/d\Ts$ term in \eqnref{beta_def}.  Indeed, the vast majority of models predict a non-negligible  $dS/d\Ts > 0$ \citep[e.g][]{donohoe2014,trenberth2009},  due primarily to increased absorption of near-infrared sunlight by water vapor,\footnote{Not to be confused with the thermal or `far' infrared light we have been discussing.}   as well as a decrease in ice cover which reduces reflection and allows for more solar absorption (the `ice-albedo' feedback). Though we could attempt to further refine our estimate by modeling these other effects, no single one of them  seems to impact climate sensitivity as strongly as the water vapor feedback \citep[][]{soden2006}, so we leave the matter here.   

\section{Why does mean precipitation increase with warming?}
\label{sec_dpdts}
Though we do not pursue simple models of climate sensitivity any further, there are other phenomena of interest besides surface temperature which vary with climate. Probably the next most important such phenomenon is precipitation. While there are many important questions one could ask about precipitation, perhaps the most basic concerns the overall strength of the hydrological cycle: how does global and annual mean precipitation $P$ (units $\kg/\meter^2/\second$) respond to increases in \Ts? Models robustly predict an increase of 2-3 \% per Kelvin \citep[e.g.][]{lambert2008,stephens2008a,held2006}, but the origin of these numbers is not widely understood. It turns out, however, that $P$ and its changes are closely tied to the radiation physics of the previous section. This section explores this connection to try and understand where this $2-3 \%\  \Kinverse$ number  comes from.

%
	\subsection{The radiative constraint on precipitation}

The connection between $P$ and infrared radation can be gleaned from Fig. \ref{rce_cartoon} and the associated discussion. If we focus on the energy budget of the \emph{atmosphere}, rather than the energy flows through the whole system, then (as pointed out in Section \ref{sec_2box}) we see that the atmospheric energy balance is between thermal infrared cooling on the one hand and latent heating from condensation of precipitation on the other.\footnote{For simplicity we assume that the atmosphere is transparent to solar radiation, even though absorption of solar radiation can be significant \citep[e.g.][]{trenberth2009a}. The arguments we make below are easily extended to solar absorption, however  \citep{jeevanjee2017c}, so this approximation does not affect the validity of our arguments.} We formalize this as follows. Let $F(z)$ denote the net (upwelling  minus downwelling)\footnote{We here employ the \emph{two-stream approximation} and assume that a fully three-dimensional radiation field can be effectively described by upwelling and downwelling fluxes. See \cite{thomas2002} for details.} thermal radiation flux at height $z$ (units \Wmsq), so that $\OLR = F(\infty)$. Then define  
\beqn
	Q\ \equiv \ F(\infty) - F(0) \ .
	\label{Q_def}
\eeqn
Since $Q$ is just the net flux out of the atmosphere $F(\infty)$ minus the net flux into the atmosphere $F_0$, it represents the the total thermal infrared cooling in the column.\footnote{Note that $F(0)$ is the residual between the upwards and downwards red arrows near the surface in Fig. \ref{rce_cartoon}. In section \ref{sec_rce} we implied that these two arrow cancelled exactly and hence that $F(0)=0$, but this is not quite true. For the present day tropics $F(0)\approx 40 \ \Wmsq$, and it increases with decreasing \Ts, so can be significantly higher elsewhere over the globe.}    Atmospheric energy balance then simply states that 
\beqn
	LP \ = \ Q \ .
	\label{rad_constraint}
\eeqn
We thus have a  \emph{radiative constraint} on precipitation, namely  that (up to a constant conversion factor of $L$) mean precipitation  should  equal  total atmospheric cooling.\footnote{This constraint is only approximate, as it ignores the direct (conduction) heating of the tropical atmosphere by the relatively warm ocean. For more on this see, e.g., \cite{ogorman2012}.}  Thus, in particular, we may try to understand changes in $P$ with surface warming via changes in $Q$.

\subsection{Flux divergence and emission to space }
	
To proceed, we note that by its definition \eqnref{Q_def}, $Q$ can be written as a vertical integral, where as in the previous section we use temperature as our vertical coordinate:
\beqn
	Q\ = \ \int_{\Ttp}^{\Ts} (-\partial_{T} F)\, dT
	\label{Q_int}
\eeqn
(we assume that $F(T=\Ttp) \approx F(z=\infty)$).	The \emph{flux divergence} $-\ppt F$ (units \WmsqK) is a vertically resolved radiative cooling, giving the watts of cooling from a layer of atmosphere with unit area and unit temperature difference across its vertical extent.  We can then \emph{spectrally} resolve this cooling  as
\beqn
	-\ppt F \ = \ \int_0^\infty (-\ppt \Flambda) \, d\lambda
\eeqn
where $\Flambda$ (units $\Wmsq/\meter$) is the  thermal radiation flux at wavelength $\lambda$, with corresponding optical depth \taulambda. The behavior of the spectrally and vertically resolved radiative cooling (flux divergence) $-\ppt \Flambda$ will be the key to understanding $Q$ and $P$.

 To  proceed, however,  we first need the basic notion of transmissivity. Dropping the $\lambda$ subscript for the moment, suppose we have an upwelling flux $U(\tau_0)$ at  optical depth $\tau_0$, and no downwelling flux at all. How will $U(\tau)$ attenuate as it travels upward, assuming no other sources? After traveling upwards through optical depth $ \Delta \tau \ll 1$, we know that absorbers have obscured a fraction $\Delta \tau$ of the total area of the column, and hence the magnitude of the change in $U$ will be (recalling that $\tau$ decreases upwards)
 \begin{align}
 	 U(\tau_0)\ - \ U(\tau_0 - \Delta \tau)\  & =\  U \Delta \tau \n \\
	\implies \frac{dU}{d\tau}\ & =\  U \n \\
	\implies U(\tau)\ & = \ U(\tau_0) e^{-(\tau_0-\tau)}  \n \\
	\implies   U(0) \ &  = \ U(\tau_0) e^{-\tau_0} \ . \label{eqn_trans}
\end{align}
Eqn. \eqnref{eqn_trans} is saying that only a fraction $e^{-\tau_0}$ of the flux emitted at $\tau_0$ escapes to space ($\tau=0$). We thus say, in general, that $e^{-\tau}$ is the \emph{transmissivity} of the atmosphere at level $\tau$. 
 
   Now consider a layer of atmosphere at temperature $T$ with differential optical thickness $d\taulambda > 0$. Its effective emitting area per unit area is $d\taulambda$, so its upward emission (per unit area) at wavelength $\lambda$ is $\pi B(\lambda, T) d\taulambda$. By \eqnref{eqn_trans}, the fraction of this emission that makes it to space is $e^{-\taulambda}$, so its emission to space (or \emph{cooling-to-space})  is $\pi B(\lambda, T)e^{-\taulambda} d\taulambda$. 
Of course, this layer will also exchange radiation with cooler layers above it, which will cool the layer, as well as exchange radiation with warmer layers below (including the surface), which will warm it. However, it turns out that so long as we're not too close to the surface, these `exchange terms' roughly cancel each other, and so the \emph{total} cooling in the layer is roughly equal to the cooling to space. Finite differencing \Flambda\ across this layer and dividing by its temperature depth $dT$ gives  
 \beqn
 		-\ppt \Flambda\ \approx \ B(\lambda, T)e^{-\taulambda}\frac{d \taulambda}{dT}  \ .
	\label{cts}
\eeqn
Equation \eqnref{cts} is known as  the \emph{cooling-to-space} approximation \citep{rodgers1966}.
 The point of writing down Eqn. \eqnref{cts} is that, together with Eqn. \eqnref{eqn_taut} and the manifest \Ts-invariance of $B(\lambda,T)$, it implies that the flux divergence   $-\ppt \Flambda$ (and hence $-\ppt F$ by spectral integration) is also \Ts-invariant.

 \subsection{\Ts-invariance  and the deepening troposphere}
 With the \Ts-invariance of $-\ppt F$  in hand, we can now gain insight into how $Q$ and $P$ change with \Ts\ \citep[following][]{jeevanjee2017c}. The first step is to note that  $-\ppt F$ is the integrand in  our expression \eqnref{Q_int} for $Q$. Furthermore, the tropopause temperature $\Ttp$ which forms the lower limit on that integral is just the temperature at which $-\ppt F$ goes to 0, and is hence itself \Ts-invariant.\footnote{In realistic simulations this claim is complicated somewhat by ultraviolet absorption by ozone in the stratosphere and a consequent `cold-point' or temperature minimum, which is slightly higher in altitude than where $-\ppt F$ goes to 0 in the tropics. This cold-point temperature actually exhibits robust temperature increases with increasing \Ts, but the increases are generally only a fraction of the \Ts\ increase; thus  \Ts-invariance is still a reasonable first approximation. See e.g. \cite{lin2017,kim2013a} and references therein.}
  (Earlier we defined $\Ttp$ as the temperature at the top of the convecting layer, but since convective heating balances radiative cooling, convection stops roughly where radiative cooling stops.) Also, $-\ppt F$ increases more or less monotonically throughout the depth of the atmosphere. 
 
\begin{figure}[t]
        \begin{center}
                \includegraphics[scale=0.5,trim = 0cm 4cm 0cm 6cm,clip=true]{\figurepath 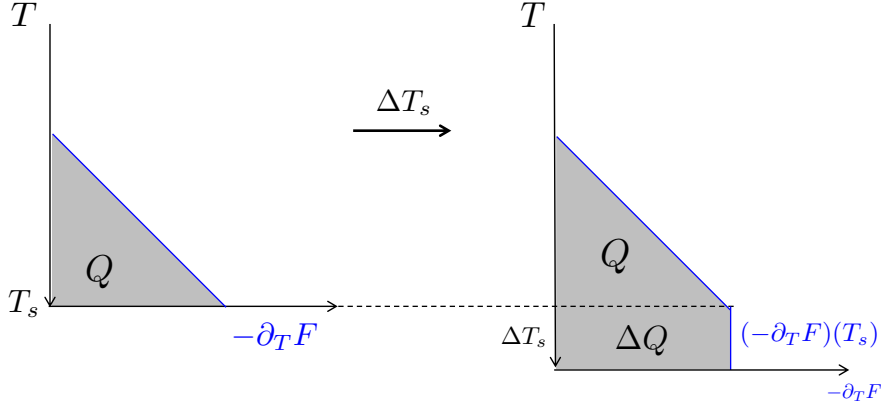}
                \caption{Cartoon of the \Ts-invariance of $(-\ppt F)(T)$ (blue lines) and the resulting picture for $dQ/d\Ts$, which arises from a deepening of the atmosphere from below when viewed in temperature coordinates. As in proofs of the fundamental theorem of calculus, we assume a constant (rather than increasing) value of $(-\ppt F)(T)$ over the interval $\Delta \Ts$; this approximation yields vanishingly small errors in the limit  $\Delta \Ts \rightarrow 0$. For realistic $(-\ppt F)(T)$ profiles see \cite{jeevanjee2017c}.
                \label{dqdts_cartoon}
                }
        \end{center}
\end{figure}

Putting these pieces together yields the picture in Fig. \ref{dqdts_cartoon}. Because \Ttp\ is \Ts-invariant,  global warming is a \emph{deepening of the atmosphere from below} when viewed in temperature coordinates. Furthermore, as far as $-\ppt F$ and $Q$ are concerned, warming simply exposes more of the \Ts-invariant $-\ppt F$ profile, which then adds to the integral in \eqnref{Q_int}, increasing $Q$ and hence $P$. We may quantify this by differentiating Eqn. \eqnref{Q_int} with respect to \Ts\ and applying the fundamental theorem of calculus, which yields
\beqn
	\frac{dQ}{d\Ts} \ = \ (-\ppt F)(\Ts).
	\label{eqn_dqdts}
\eeqn
 For a tropical atmosphere at \Ts=300 K, and for the moment including atmospheric  absorption of sunlight, we find $Q\approx 100 \ \Wmsq$ and $(-\ppt F)(\Ts) \approx 3 \ \WmsqK $ \citep{jeevanjee2017c}, yielding $d\ln Q/d\Ts = 3 \%\ \Kinverse$, consistent with global warming simulations.
 
 While Eqn. \eqnref{eqn_dqdts} gives us a way to predict $dQ/d\Ts$ without running a global warming simulation, it does not necessarily provide much insight into \emph{why} $d\ln Q/d\Ts = 2-3 \%\ \Kinverse$. For this, let us build an even simpler model, wherein we parameterize $-\ppt F \sim (T-\Ttp)^\beta$, where realistic profiles of $-\ppt F$ suggest $\beta = 1-2$ \citep{jeevanjee2017c}. Then we have
 \begin{align}
 	Q  &\ \sim \ (\Ts - \Ttp)^{\beta + 1}  \n \\
	\implies \frac{d \ln Q}{d \Ts} & \  = \frac{\beta + 1}{\Ts - \Ttp} \ = \  2-3 \% \ \Kinverse \ .  \label{eqn_dlnqdts}
 \end{align}
 (In the last step we used typical tropical values  $\Ts =300 \ \Kelvin$ and $\Ttp = 200 \ \Kelvin$.) Equation  \eqnref{eqn_dlnqdts} tells us that it is the \emph{depth of the troposphere}, $\Ts - \Ttp$, that really matters for $d \ln Q/d\Ts$. Indeed, if $-\ppt F$ were constant throughout the troposphere, then $Q$ would scale directly with $\Ts - \Ttp$, and since a 1 K increase in \Ts\ is a 1\% increase in $\Ts - \Ttp$, we would find $d \ln Q/d\Ts = 1\%\ \Kinverse$ (as predicted by \eqnref{eqn_dlnqdts} with $\beta=0$). Thus, the fact that $Q$ (and hence $P$) increase at a rate of $O(1\%)\ \Kinverse$ comes from the fact that the present-day troposphere is roughly 100 K deep. The fact that $d \ln Q/d\Ts = 2-3 \%\ \Kinverse$ rather than $1\%\ \Kinverse$ comes from the fact that $\beta \approx 1-2$ instead of 0.


\section{On constant relative humidity}
\label{sec_rh}
So far we have explored the sensitivities of \OLR\ and $P$ to surface temperature \Ts. Another climate variable of interest is relative humidity \RH. Relative humidity is directly sensible and relevant to both plant and animal life, and also played a role in the \Ts-invariance discussed in the previous sections, where it was assumed implicitly that \RH\ does not vary significantly with \Ts.  What is the basis for this? More generally, how do we expect \RH\ to change with \Ts?
 
 To proceed we must first distinguish between \RH\ in the boundary layer (Fig. \ref{rce_cartoon}, $z \lesssim 1\ \km$) and the so-called `free troposphere'  ($ 1 \lesssim z \lesssim 15 \ \km$), where in the latter there is moist convection, precipitation, and latent heat release. The sources and sinks of water vapor, as well as their physics, is thus quite different in these two regions, and they must be treated separately.

\subsection{Boundary layer relative humidity} 
 
 We begin with boundary layer relative humidity \RHbl, following a standard argument.\footnote{Articulated, for instance, in https://www.gfdl.noaa.gov/blog\_held/47-relative-humidity-over-the-oceans/.} It turns out the dynamics of \RHbl\ are heavily influenced by the dynamics of $Q$ and $P$, which we discussed in the previous section. The key link here is the evaporation rate $E$ (units $\kg/\meter^2/\sec$), which in a steady state  must obey  
 \beqn
 E\ = \ P \ 
\label{P=E}
\eeqn
by conservation of water mass  (if this were not true then the atmosphere would be continuously moistening or drying, and thus not be in steady state).  Furthermore, $E$ is also controlled by \RHbl; if the boundary layer is dry then evaporation increases, and if the boundary layer is moist then evaporation decreases (that humidity inhibits evaporation is evident from humid summer days). Stronger wind speeds \speed\ and turbulence should also increase $E$ (as we know from experiencing gusts of wind after exiting a pool), as they will more efficiently replace the recently moistened air right above the surface with unmoistened air. All of this is  formalized in the \emph{bulk aerodynamic formula} for evaporation:
 \begin{subequations}
	 \begin{align}
 		E \ & =\   \Cd \speed \left[\rhovstar(\Ts) - \rhov\right] \label{bulk_E1} \\
	  		&  \approx \ \Cd \speed \rhovstar(\Ts)\,(1-\RHbl) \ . \label{bulk_E2}
	\end{align}
\end{subequations}
 It is assumed in \eqnref{bulk_E1} that \speed\ and the BL vapor density \rhov\ are evaluated at some reference height \zref\ near the surface, often taken to be 6 m. The quantity $\rhovstar(\Ts)$ is the \emph{saturation vapor density}  [i.e. \eqnref{eqn_rhov} with \RH=1] at the surface temperature \Ts, and 
represents the surface moisture. If we assume that $T(\zref) \approx \Ts$ (a good approximation throughout the tropics, where the difference is roughly 1 K) then we can write $\rhov\approx \RHbl\rhovstar(\Ts)$, as in \eqnref{bulk_E2}. The factor $\Cd \sim 10^{-3}$ is a dimensionless `drag coefficient' parameterizing the roughness of the surface and the resulting turbulence.
The form of \eqnref{bulk_E1} can be deduced from general principles, but some empiricism is needed to determine \Cd\     \citep[see][and references therein]{hartmann2015book,pierrehumbert2010}.

We are now in a position to estimate how  \RHbl \ changes with \Ts. We do this by calculating the logarithmic derivative of \eqnref{bulk_E2} with respect to \Ts, as follows. On the left-hand side we replace $E$ with $P$, knowing from Section \ref{sec_dpdts} that $d \ln P/d\Ts \approx 0.02 \ \Kinverse$. On the right, we assume that \Cd\ and \speed\ don't change appreciably with warming,\footnote{An approximation borne out by models; see, e.g.,  \cite{laine2014}.}  so all that's left are the  $\rhovstar(\Ts)$ and  $(1-\RHbl)$ factors. For the former, we note that by \eqnref{eqn_rhov} the dominant $T$-dependence is in $p_v^*(\Ts)$, and from \eqnref{CC} we see that its logarithmic derivative is 
\beqn
 \frac{d \ln p_v^*}{d \Ts} \ = \ \frac{L}{R_v \Ts^2}  \ \approx \ 0.06  \ \Kinverse
 \label{CC_scaling}
 \eeqn
 at $\Ts= 300\ \Kelvin$. Putting these ingredients together,  rearranging,  and assuming a typical tropical $\RHbl \approx 0.75$, we find
 \beqn
 	\frac{d\, \RHbl}{d\Ts}\ = \ \left(\frac{L}{R_v\Ts^2}  \ - \ \frac{d \ln P}{d \Ts}\right)(1-\RHbl) \ \approx \ (0.04\ \Kinverse)(1/4) \ = \ 0.01 \ \Kinverse \ ,
	\label{drhdts}
\eeqn
an estimate consistent with model results \citep{laine2014}. Thus \RHbl\ is not quite constant with respect to \Ts, but its sensitivity is still quite small: equation \eqnref{drhdts} implies that under a doubling of \cotwo\ with  $\ECS \approx 3.6 \ \Kelvin$, one might expect tropical \RHbl\ to increase by only about 0.035.  Equation \eqnref{drhdts} shows that this is because $d\RHbl/d\Ts$ is reduced relative to the logarithmic sensitivities of $\rhovstar$ and $P$ to \Ts\ by a factor $(1-\RHbl) \approx 1/4$.

The chain of reasoning here may seem circular: we assumed small changes in \RH\ to deduce $d \ln P/d\Ts \approx 0.02 \ \Kinverse$, but then used the latter to deduce that changes in \RHbl\  are small. But what we really have are constraints on  $d \ln P/d\Ts$ and $d \RHbl/d\Ts$ [Eqns.  \eqnref{eqn_dqdts} and \eqnref{drhdts}, in conjunction with \eqnref{rad_constraint}] which are mutually consistent, and the nontrivial finding is that Eqn. \eqnref{drhdts} predicts a small value of $d \RHbl/d\Ts$, consistent with the value of $d \ln P/d\Ts$ which we plugged in.

\subsection{Free tropospheric relative humidity}
\label{sec_RHft}
In the free troposphere, relative humidity \RHft\ is set by different processes. Here the source of water vapor at a given height $z$ is `detrainment' of saturated air from rising convecting clouds: as a convecting parcel rises, it mixes with its environment and leaves some of its saturated air behind, moistening the environment. The free tropospheric \emph{sink} of water vapor derives from the opposing motion of the environment: as clouds rise environmental air must sink, and so at any given height, air from above (which is colder and hence drier) moves downwards, drying the environment at that height. The free tropospheric relative humidity is set by a balance between these two processes. 

To formalize this we need a model of convection which goes beyond simply relating surface and atmospheric temperatures (as  in Section \ref{sec_rce}) and somehow models how convection mixes the troposphere.  A simple model which fits the bill is the `bulk-plume' model of convection, which we employ following the treatment in \cite{romps2014}. For simplicity we assume throughout that all condensed water in clouds precipitates out immediately; the effects of this simplification on the bulk-plume model and the moist adiabatic lapse rate derived below are discussed in \cite{romps2014} and \cite{emanuel1994book}, respectively. 

\begin{figure}[t]
        \begin{center}
                \includegraphics[scale=0.4,trim = 0cm 0cm 0cm 0cm,clip=true]{\figurepath 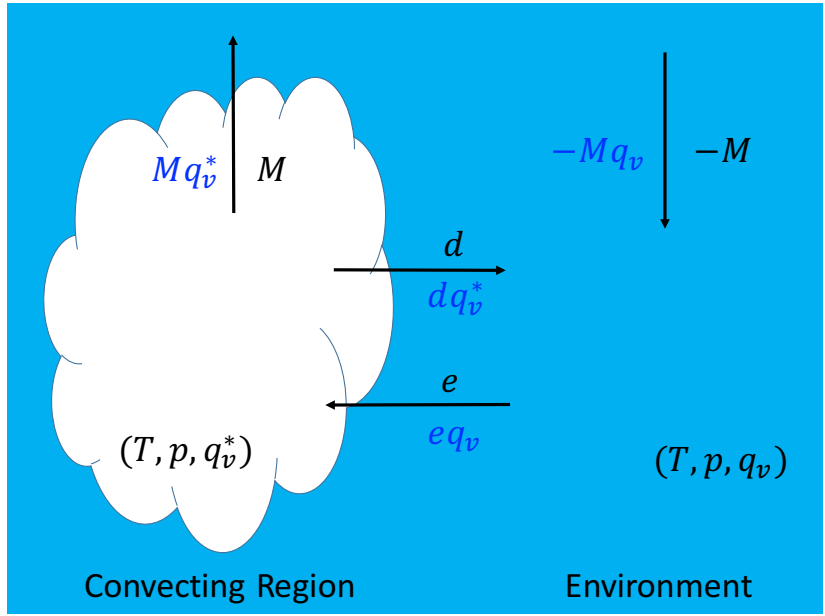}
                \caption{Cartoon of the bulk-plume model of convection. Mass fluxes, sources, and sinks are in black whereas those for moisture are in blue. The single cloud on the left should be thought of as comprising  multiple convective plumes which may not be spatially contiguous but which all have the same thermodynamic properties at a given height.
                \label{bulk_plume_cartoon}
                }
        \end{center}
\end{figure}

The bulk-plume model  partitions the atmosphere into a cloudy, convecting `region'  (which we should think of as comprising  multiple convective plumes which may not be spatially contiguous)  and a clear environment, where both regions are assumed to be horizontally homogenous (i.e. all the convective plumes have the same properties at a given height; this is the meaning of `bulk' in `bulk-plume'). See Figure \ref{bulk_plume_cartoon}. The convecting region is characterized by a convective mass flux $M(z)$ $(\kg/\meter^2/\second)$, which represents the kilograms of convecting air passing upwards through an \emph{average} square meter of the total domain per second. Note that essentially by conservation of mass  there must be a compensating, negative, `subsidence' mass flux $-M$ in the environment; in other words, clear air sinks.\footnote{Though the mass fluxes in the two regions are equal and opposite, the same is not true of their vertical velocities; in fact, the subsidence vertical velocity is $O(10^{-3})\ \meter/\second$ whereas that in the convecting region is $O(1)\ \meter/\second$. This is compensated for by the fact that the areal coverage of the convecting region is roughly 1000 times smaller than that of the environment.} Since mixing between the two regions is necessary for clouds to moisten the environment and hence for $\RHft\neq 0$, we assume that environmental air is mixed \emph{in} to the convecting region at a rate $e(z)$  and that cloudy air is mixed \emph{out} of the convecting region at a rate $d(z)$ (Fig. \ref{bulk_plume_cartoon}). These are known as the \emph{entrainment} and \emph{detrainment} rates, respectively (units $\kg/\meter^3/\second$, also averaged over the whole domain). They  are both positive definite, and satisfy
\beqn
	\ddz M  \ = \ e  - d \ .
\label{bulk_M}
\eeqn

Equation \eqnref{bulk_M} describes how the total convective mass flux $M(z)$ evolves with height due to the source and sink profiles $e(z)$ and $d(z)$. To proceed further we need an analogous equation for the total \emph{moisture} flux. The most convenient way to do this is to change moisture variables from the vapor density  \rhov\ of the previous sections to the \emph{specific humidity}
\beqn
 \qv \ \equiv \ \frac{\rhov}{\rho}\ .  
 \label{qv_def}
 \eeqn

We do this because our convective (environmental) moisture flux  is then simply $M$ (-$M$) times the corresponding $\qv$. But how are the \qv\ in the convecting region and environment related? At a given height we can assume that $p$ is equal in both regions, but what about temperature? It turns out that the temperature of convecting parcels, while higher than that of the environment to provide buoyancy, are only slightly higher \citep[order 1 K or less, e.g.][]{romps2015stereo}, 
so we assume that $T$ is equal in both regions as well.  Hence, the main difference between the two regions is in their water content, i.e. that the convecting region is saturated (by assumption). Combining \eqnref{qv_def} with  \eqnref{eqn_rhov} and the ideal gas law \eqnref{ideal_gas} yields an expression for this saturation specific humidity:
\beqn
	\qvstar  \ = \ \frac{ R_d\pvstar(T)}{R_v p} \ .
	\label{qvstar}
\eeqn
 
%
Note that \qvstar\ is a function of $T$ and $p$. We denote the environmental specific humidity as simple \qv. With this, the source of environmental moisture\footnote{We neglect here the source of water vapor from evaporating condensates, i.e. we assume that all water that condenses falls instantly to the ground as precipitation. We also omit here the moisture equation for the convecting region. These additions and much more can be found in \cite{romps2014}.}  is $d\qvstar$ and the sink is $e\qv$, and thus the environmental moisture flux  is governed by
\beqn
\ddz (-M\qv) \ = \ d\qvstar - e\qv \ . 
\label{bulk_Mq1}
\eeqn
At this point it is convenient to introduce the \emph{fractional} entrainment and detrainment rates $\epsilon \equiv e/M$ and $\delta \equiv d/M$ (units $1/\meter$), which just give the fraction of a parcel's mass which is entrained or detrained per unit height of ascent. Using these, substituting \eqnref{bulk_M} into \eqnref{bulk_Mq1}, and noting that $\RHft = \qv/\qvstar$, we obtain
\beqn
	-\ddz (\RHft \qvstar) \ = \ \delta \qvstar(1-\RHft) \ .
\label{bulk_Mq2}
\eeqn
Assuming that $\ddz \ln \RHft \ll \ddz \ln \qvstar$ in magnitude, setting
\beqn
	\gamma \equiv -  \frac{d \ln \qvstar}{dz} 
	\label{gamma_def}
\eeqn 
(i.e. $\gamma$ is the fractional change in \qvstar\ with height),  and rearranging \eqnref{bulk_Mq2} then yields our desired expression for free-tropospheric relative humidity,
\beqn
	\RHft \ = \ \frac{\delta}{\delta + \gamma} \ .
	\label{eqn_rh}
\eeqn

What does this equation tell us? First off, note that when convective mixing is vigorous ($\delta \gg \gamma$), Eqn. \eqnref{eqn_rh} gives $\RH \approx 1$, as expected. Conversely, in the opposite limit of weak convective mixing ($\delta \ll \gamma$), Eqn. \eqnref{eqn_rh} gives $\RH \approx 0$, also as expected. Of course, realistic values of $\delta$ and $\gamma$ don't lie in either limit, and the typical values\footnote{This value of $\delta$ is taken from simulations, e.g.  \cite{romps2014rayleigh,boing2012b}. As for $\gamma$, this value comes from setting $(T,p) =(255 \ \Kelvin,\ 0.5\ \text{atm})$ in \eqnref{gamma_m} and \eqnref{gamma_exp}.}  $\delta = 1\ \km^{-1}$  and $\gamma = 0.5 \ \km^{-1}$  yield $\RHft \approx 0.67$, a reasonable mid-tropospheric value \citep{romps2014}. 

How are these values expected to change with surface warming? The sensitivity of $\delta$ to \Ts\ is relatively unexplored and needs further work, but we do know that $\Ttp$ is \Ts-invariant, and that convection stops at \Ttp. Thus detrainment rates must peak near \Ttp\ regardless of \Ts, giving $\delta$ profiles at least some degree of \Ts-invariance. As for $\gamma$, we may substitute \eqnref{qvstar} into \eqnref{gamma_def} to obtain
\beqn
	\gamma \ = \ \frac{L\Gamma_m}{R_vT^2} - \frac{g}{R_dT} \ .
	\label{gamma_exp}
\eeqn
We can obtain an explicit expression for $\Gamma_m$ by following the derivation of \eqnref{dry_lapse} but now including a diabatic heating term from condensation, $dQ = -\rho V L d\qvstar$. Performing the thermodynamic manipulation\footnote{This is most easily done by starting with the identity $dQ = \rho V C_p dT - V dp$ and writing each term in terms of $dz$. After cancellation of the common $\rho V dz$ factors, this yields
\beqn
	 L \qvstar \gamma \ = \  -C_p\Gamma_m + g \ .
	 \n
\eeqn
 Plugging in \eqnref{gamma_exp} and solving for $\Gamma_m$ then yields \eqnref{gamma_m}.}
then yields
\beqn
	\Gamma_m \ = \ \frac{g}{C_p}\frac{1 + \frac{L\qvstar}{R_d T}}{1 + \frac{L^2\qvstar}{C_pR_vT^2} } \ ,
\label{gamma_m}
\eeqn
a moist generalization of \eqnref{dry_lapse}. Since \qvstar\ is exponential in $T$ (via $p_v^*(T)$) it dominates variations in $\Gamma_m$.\footnote{Indeed, as $(T,p)$ vary from $(300 \ \Kelvin, \ 1000 \ \mathrm{hPa})$ at the surface to roughly $(200\  \Kelvin, \ 200\  \mathrm{hPa})$ near the tropopause, \qvstar\ drops from 0.02 to $10^{-5}$, yielding $\Gamma_m \approx 4 \ \Kelvin/\km$ at the surface and $\Gamma_m \approx \Gamma_d = 10 \ \Kelvin/\km$ near the tropopause, as claimed in Section \ref{sec_rce}.} Furthermore, for regimes of interest the $p$-dependence of \qvstar\ is small \citep{romps2014}. This implies, then, that  $\Gamma_m(T)$  is  approximately \Ts-invariant, as claimed in Section \ref{sec_h2o_feedback} \citep[see also Fig. 2 of][]{ingram2010}. Since $T$ is the only other variable entering the rest of \eqnref{gamma_exp}, we see that $\gamma(T)$ itself should be approximately \Ts-invariant.  

All of this suggests that $\RHft(T)$ should be \Ts-invariant, similar to what we found for $(-\ppt F)(T)$, and cloud-resolving  simulations indeed confirm this \citep{romps2014}. Given that gradients in $\RHft(T)$ are typically less than $0.01\ \Kinverse$, we would also expect that at fixed $z$ or $p$, changes in \RHft\ per degree \Ts\ should be $O(0.01)$, and  this is consistent with  GCM results \citep{sherwood2010}. Such small changes in \RHft\ should then lead to a water vapor feedback roughly consistent with constant \RHft, as claimed here and also as found in GCMs \citep[e.g.][]{soden2006}. The arguments given here give one way of understanding these results.

\bibliographystyle{apa}

\end{document}